# QUANTUM MECHANICS IN THE LIGHT OF QUANTUM COSMOLOGY

**Murray Gell-Mann**
*California Institute of Technology*
*Pasadena, CA 91125   USA*

and

**James B. Hartle**
*Department of Physics*
*University of California*
*Santa Barbara, CA 93106   USA*

## ABSTRACT

We sketch a quantum-mechanical framework for the universe as a whole. Within that framework we propose a program for describing the ultimate origin in quantum cosmology of the "quasiclassical domain" of familiar experience and for characterizing the process of measurement. Predictions in quantum mechanics are made from probabilities for sets of alternative histories. Probabilities (approximately obeying the rules of probability theory) can be assigned only to sets of histories that approximately decohere. Decoherence is defined and the mechanism of decoherence is reviewed. Decoherence requires a sufficiently coarse-grained description of alternative histories of the universe. A quasiclassical domain consists of a branching set of alternative decohering histories, described by a coarse graining that is, in an appropriate sense, maximally refined consistent with decoherence, with individual branches that exhibit a high level of classical correlation in time. We pose the problem of making these notions precise and quantitative. A quasiclassical domain is emergent in the universe as a consequence of the initial condition and the action function of the elementary particles. It is an important question whether all the quasiclassical domains are roughly equivalent or whether there are various essentially inequivalent ones. A measurement is a correlation with variables in a quasiclassical domain. An "observer" (or information gathering and utilizing system) is a complex adaptive system that has evolved to exploit the relative predictability of a quasiclassical domain, or rather a set of such domains among which it cannot discriminate because of its own very coarse graining. We suggest that resolution of many of the problems of interpretation presented by quantum mechanics is to be accomplished, not by further scrutiny of the subject as it applies to reproducible laboratory situations, but rather by an examination of alternative histories of the universe, stemming from its initial condition, and a study of the problem of quasiclassical domains.

*This paper was the first in a series by the two authors developing a quantum mechanical framework for the universe as a whole called Decoherent Histories Quantum Mechanics, DH. Although developed independently, and differing on certain points, DH has much in common with with Robert Griffiths' and Roland Omnès' earlier consistent histories formulation of quantum mechanics as explicitly acknowledged. The paper has not been updated or improved and the references are unchanged. The paper appeared in the* Proceedings of the Santa Fe Institute Workshop on Complexity, Entropy, and the Physics of Information, *May 1989 and in the* Proceedings of the 3rd International Symposium on The Foundations of Quantum Mechanics in the Light of New Technology, *Tokyo, Japan, August 1989. It is posted here to make it accessible to those who might not have easy access to the published sources.*

# I. QUANTUM COSMOLOGY

If quantum mechanics is the underlying framework of the laws of physics, then there must be a description of the universe as a whole and everything in it in quantum-mechanical terms. In such a description, three forms of information are needed to make predictions about the universe. These are the action function of the elementary particles, the initial quantum state of the universe, and, since quantum mechanics is an inherently probabilistic theory, the information available about our specific history. These are sufficient for every prediction in science, and there are no predictions that do not, at a fundamental level, involve all three forms of information.

A unified theory of the dynamics of the basic fields has long been a goal of elementary particle physics and may now be within reach. The equally fundamental, equally necessary search for a theory of the initial state of the universe is the objective of the discipline of quantum cosmology. These may even be related goals; a single action function may describe both the Hamiltonian and the initial state.[a]

There has recently been much promising progress in the search for a theory of the quantum initial condition of the universe.[3] Such diverse observations as the large scale homogeneity and isotropy of the universe, its approximate spatial flatness, the spectrum of density fluctuations from which the galaxies grew, the thermodynamic arrow of time, and the existence of classical spacetime may find a unified, compressed explanation in a particular simple law of the initial condition.

The regularities exploited by the environmental sciences such as astronomy, geology, and biology must ultimately be traceable to the simplicity of the initial condition. Those regularities concern specific individual objects and not just reproducible situations involving identical particles, atoms, etc. The fact that the discovery of a bird in the forest or a fossil in a cliff or a coin in a ruin implies the likelihood of discovering another similar bird or fossil or coin cannot be derivable from the laws of elementary particle physics alone; it must involve correlations that stem from the initial condition.

The environmental sciences are not only strongly affected by the initial condition but are also heavily dependent on the outcomes of quantum-probabilistic events during the history of the universe. The statistical results of, say, proton-proton scattering in the laboratory are much less dependent on such outcomes. However, during the last few years there has been increasing speculation that, even in a unified fundamental theory, free of dimensionless parameters, some of the observable characteristics of the elementary particle system may be quantum-probabilistic, with a probability distribution that can depend on the initial condition.[4]

It is not our purpose in this article to review all these developments in quantum cosmology.[3] Rather, we will discuss the implications of quantum cosmology for one of the subjects of this conference — the interpretation of quantum mechanics.

# II. PROBABILITY

Even apart from quantum mechanics, there is no certainty in this world; therefore physics deals in probabilities. In classical physics probabilities result from ignorance; in quantum mechanics they are fundamental as well. In the last analysis, even when treating ensembles statistically, we are concerned with the probabilities of particular events. We then deal the probabilities of deviations from the expected behavior of the ensemble caused by fluctuations.

When the probabilities of particular events are sufficiently close to 0 or 1, we make a definite prediction. The criterion for "sufficiently close to 0 or 1" depends on the use to which the probabilities are put. Consider, for example, a prediction on the basis of present astronomical observations that the sun will come up tomorrow at 5:59 AM ± 1 min. Of course, there is no certainty that the sun will come up at this time. There might have been a significant error in the astronomical observations or the subsequent calculations using them; there might be a non-classical fluctuation in the earth's rotation rate or there might be a collision with a neutron star now racing across the galaxy at near light speed. The prediction is the same as estimating the probabilities of these alternatives as low. How low do they have to be before one sleeps peacefully tonight rather than anxiously awaiting the dawn? The probabilities predicted by the laws of physics and the statistics of errors are generally agreed to be low enough!

All predictions in science are, most honestly and most generally, the probabilistic predictions of the *time histories* of particular events in the universe. In cosmology we are necessarily concerned with probabilities for the single system that is the universe as a whole. Where the universe presents us effectively with an ensemble of identical subsystems, as in experimental situations common in physics and chemistry, the probabilities for the ensemble as a whole yield definite predictions for the statistics of identical observations. Thus, statistical probabilities can be derived, in appropriate situations, from probabilities for the universe as a whole.[5]

Probabilities for histories need be assigned by physical theory only to the accuracy to which they are used. Thus, it is the same to us for all practical purposes whether physics claims the probability of the sun not coming up tomorrow is $10^{-10^{57}}$ or $10^{-10^{27}}$, as long as it is very small. We can therefore conveniently consider *approximate probabilities*, which need obey the rules of the probability calculus only up to some standard of accuracy sufficient for all practical purposes. In quantum mechanics, as we shall see, it is likely that only by this means can probabilities be assigned to interesting histories at all.

### III. HISTORICAL REMARKS

In quantum mechanics not every history can be assigned a probability. Nowhere is this more clearly illustrated than in the two-slit experiment (Figure 1). In the usual discussion, if we have not measured which slit the electron passed through on its way to being detected at the screen, then we are not permitted to assign probabilities to these alternative histories. It would be inconsistent to do so since the correct probability sum rules would not be satisfied. Because of interference, the probability to arrive at $y$ is not the sum of the probabilities to arrive at $y$ going through the upper and the lower slit:

$$p(y) \neq p_U(y) + p_L(y) \tag{1}$$

because

$$|\psi_L(y) + \psi_U(y)|^2 \neq |\psi_L(y)|^2 + |\psi_U(y)|^2 \quad . \tag{2}$$

If we *have* measured which slit the electron went through, then the interference is destroyed, the sum rule obeyed, and we *can* meaningfully assign probabilities to these alternative histories.

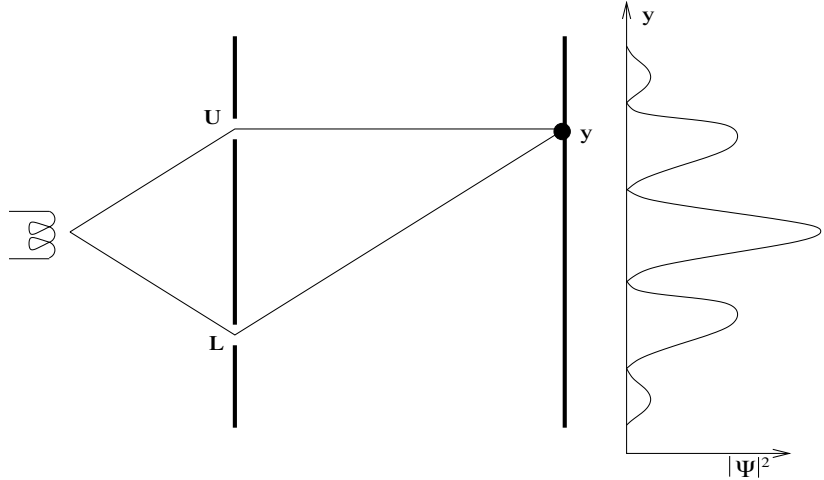

Figure 1: The two-slit experiment. An electron gun at left emits an electron traveling towards a screen with two slits, its progress in space recapitulating its evolution in time. When precise detections are made of an ensemble of such electrons at the screen it is not possible, because of interference, to assign a probability to the alternatives of whether an individual electron went through the upper slit or the lower slit. However, if the electron interacts with apparatus which measures which slit it passed through, then these alternatives decohere and probabilities can be assigned.

It is a general feature of quantum mechanics that one needs a rule to determine which histories can be assigned probabilities. The familiar rule of the "Copenhagen" interpretations described above is external to the framework of wave function and Schrödinger equation. Characteristically these interpretations, in one way or another, assumed as fundamental the existence of the classical domain we see all about us. Bohr spoke of phenomena that could be described in terms of classical language.[6] Landau and Lifshitz formulated quantum mechanics in terms of a separate classical physics.[7] Heisenberg and others stressed the central role of an external, essentially classical observer.[8] A measurement occurred through contact with this classical domain. Measurements determined what could be spoken about.

Such interpretations are inadequate for cosmology. In a theory of the whole thing there can be no fundamental division into observer and observed. Measurements and observers cannot be fundamental notions in a theory that seeks to discuss the early universe when neither existed. There is no reason in general for a classical domain to be fundamental or external in a basic formulation of quantum mechanics.

It was Everett who in 1957 first suggested how to generalize the Copenhagen framework so as to apply quantum mechanics to cosmology.[b] His idea was to take quantum mechanics seriously and apply it to the universe as a whole. He showed how an observer could be considered part of this system and how its activities — measuring, recording, and calculating probabilities — could be described in quantum mechanics.

Yet the Everett analysis was not complete. It did not adequately explain the origin of the classical domain or the meaning of the "branching" that replaced the notion of measurement. It was a theory of "many worlds" (what we would rather call "many histories"), but it did not sufficiently explain how these were defined or how they arose. Also, Everett's discussion suggests that a probability formula is somehow not needed in quantum mechanics, even though a "measure" is introduced that, in the end, amounts to the same thing.

Here we shall briefly sketch a program aiming at a coherent formulation of quantum mechanics for science as a whole, including cosmology as well as the environmental sciences.[17] It is an attempt at extension, clarification, and completion of the Everett interpretation. It builds on many aspects of the post-Everett developments, especially the work of Zeh,[18] Żurek,[19] and Joos and Zeh.[20] In the discussion of history and at other points it is consistent with the insightful work (independent of ours) of Griffiths[21] and Omnès.[22] Our research is not complete, but we sketch, in this report on its status, how it might become so.

## IV. DECOHERENT SETS OF HISTORIES

### (a) A Caveat

We shall now describe the rules that specify which histories may be assigned probabilities and what these probabilities are. To keep the discussion manageable we make one important simplifying approximation. We neglect gross quantum variations in the structure of spacetime. This approximation, excellent for times later than $10^{-43}$ sec after the beginning, permits us to use any of the familiar formulations of quantum mechanics with a preferred time. Since histories are our concern, we shall often use Feynman's sum-over-histories formulation of quantum mechanics with histories specified as functions of this time. Since the Hamiltonian formulation of quantum mechanics is in some ways more flexible, we shall use it also, with its apparatus of Hilbert space, states, Hamiltonian, and other operators. We shall indicate the equivalence between the two, always possible in this approximation.

The approximation of a fixed background spacetime breaks down in the early universe. There, a yet more fundamental sum-over histories framework of quantum mechanics may be necessary.[23] In such a framework the notions of state, operators, and Hamiltonian may be approximate features appropriate to the universe after the Planck era, for particular initial conditions that imply an approximately fixed background spacetime there. A discussion of quantum spacetime is essential for any detailed theory of the initial condition, but when, as here, this condition is not spelled out in detail and we are treating events after the Planck era, the familiar formulation of quantum mechanics is an adequate approximation.

The interpretation of quantum mechanics that we shall describe in connection with cosmology can, of course, also apply to any strictly closed sub-system of the universe provided its initial density matrix is known. However, strictly closed sub-systems of any size are not easily realized in the universe. Even slight interactions, such as those of a planet with the cosmic background radiation, can be important for the quantum mechanics of a system, as we shall see. Further, it would be extraordinarily difficult to prepare precisely the initial density matrix of any sizeable system so as to get rid of the dependence on the density matrix of the universe. In fact, even those large systems that are approximately isolated today inherit many important features of their effective density matrix from the initial condition of the universe.

### (b) Histories

The three forms of information necessary for prediction in quantum cosmology are represented in the Heisenberg picture as follows[24]: The quantum state of the universe is described by a density matrix $\rho$. Observables describing specific information are represented by operators $\mathcal{O}(t)$. For simplicity, but without loss of generality, we shall focus on non-"fuzzy", "yes-no" observables. These are represented in the Heisenberg picture by projection operators $P(t)$. The Hamiltonian, which is the remaining form of information, describes evolution by relating the operators corresponding to the same question at different times through

$$P(t) = e^{iHt/\hbar} P(0) e^{-iHt/\hbar} \quad . \tag{3}$$

An exhaustive set of "yes-no" alternatives at one time is represented in the Heisenberg picture by *sets* of projection operators $(P_1^k(t), P_2^k(t), \cdots)$. In $P_\alpha^k(t)$, $k$ labels the set, $\alpha$ the particular alternative, and $t$ its time. A exhaustive set of exclusive alternatives satisfies

$$\sum_\alpha P_\alpha^k(t) = 1 \quad , \quad P_\alpha^k(t) P_\beta^k(t) = \delta_{\alpha\beta} P_\alpha^k(t) \quad . \tag{4}$$

For example, one such exhaustive set would specify whether a field at a point on a surface of constant $t$ is in one or another of a set of ranges exhausting all possible values. The projections are simply the projections onto eigenstates of the field at that point with values in these ranges. We should emphasize that an exhaustive set of projections need not involve a *complete* set of variables for the universe (one-dimensional projections) — in fact, the projections we deal with as observers of the universe typically involve only an infinitesimal fraction of a complete set.

Sets of alternative histories consist of *time sequences* of exhaustive sets of alternatives. A *history* is a particular sequence of alternatives, abbreviated $[P_\alpha] = (P_{\alpha_1}^1(t_1), P_{\alpha_2}^2(t_2), \cdots, P_{\alpha_n}^n(t_n))$. A *completely fine-grained* history is specified by giving the values of a complete set of operators at all times. One history is a *coarse graining* of another if the set $[P_\alpha]$ of the first history consists of sums of the $[P_\alpha]$ of the second history. The inverse relation is fine graining. The completely coarse-grained history is one with no projections whatever, just the unit operator!

The reciprocal relationships of coarse and fine graining evidently constitute only a partial ordering of sets of alternative histories. Two arbitrary sets need not be related to each other by coarse/fine graining. The partial ordering is represented schematically in Fig. 2, where each point stands for a set of alternative histories.

Feynman's sum-over-histories formulation of quantum mechanics begins by specifying the amplitude for a completely fine-grained history in a particular basis of generalized coördinates $Q^i(t)$, say all fundamental field variables at all points in space. This amplitude is proportional to

$$\exp(iS[Q^i(t)]/\hbar), \tag{5}$$

where $S$ is the action functional that yields the Hamiltonian, $H$. When we employ this formulation of quantum mechanics, we shall introduce the simplification of ignoring fields with spins higher than zero, so as to avoid the complications of gauge groups and of fermion fields (for which it is inappropriate to discuss eigenstates of the field variables.) The operators $Q^i(t)$ are thus various scalar fields at different points of space.

Let us now specialize our discussion of histories to the generalized coördinate bases $Q^i(t)$ of the Feynman approach. Later we shall discuss the necessary generalization to the case of an arbitrary basis at each time $t$, utilizing quantum-mechanical tranformation theory.

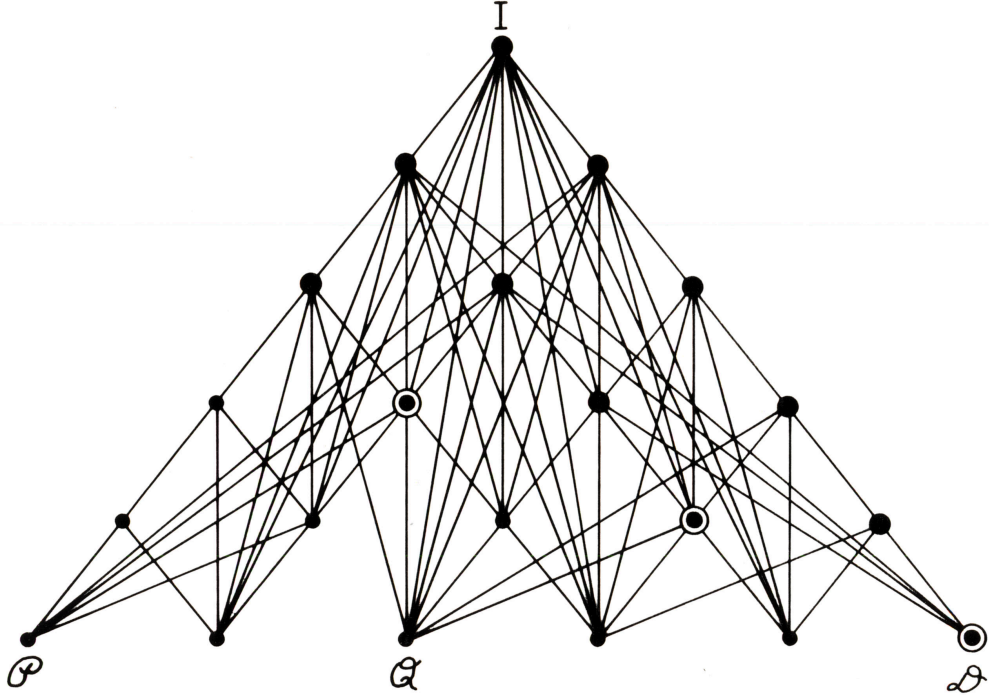

Figure 2. The schematic structure of the space of *sets* of possible histories for the universe. Each dot in this diagram represents an exhaustive *set* of alternative histories for the universe. (This is not a picture of the branches defined by a given set!) Such sets, denoted by $\{[P_\alpha]\}$ in the text, correspond in the Heisenberg picture to time sequences $(P^1_{\alpha_1}(t_1),\ P^2_{\alpha_2}(t_2),\ \cdots\ P^n_{\alpha_n}(t_n))$ of sets of projection operators, such that at each time $t_k$ the alternatives $\alpha_k$ are an orthogonal and exhaustive set of possibilities for the universe. At the bottom of the diagram are the completely fine-grained sets of histories each arising from taking projections onto eigenstates of a *complete set* of observables for the universe at *every time.* For example, the set $\mathcal{Q}$ is the set in which all field variables at all points of space are specified at every time. This set is the starting point for Feynman's sum-over-histories formulation of quantum mechanics. $\mathcal{P}$ might be the completely fine-grained set in which all field momenta are specified at each time. $\mathcal{D}$ might be a degenerate set of the kind discussed in Section VII in which the *same* complete set of *operators* occurs at every time. But there are many other completely fine-grained sets of histories corresponding to all possible combinations of complete sets of observables that can be taken at every time.

The dots above the bottom row are coarse-grained sets of alternative histories. If two dots are connected by a path, the one above is a coarse graining of the one below — that is, the projections in the set above are *sums* of those in the set below. A line, therefore, corresponds to an operation of coarse graining. At the very top is the degenerate case in which complete sums are taken at every time, yielding no projections at all other than the unit operator! The space of sets of alternative histories is thus partially ordered by the operation of coarse graining.

The heavy dots denote the decoherent sets of alternative histories. Coarse grainings of decoherent sets remain decoherent. Maximal sets, the heavy dots surrounded by circles, are those decohering sets for which there is no finer-grained decoherent set.

Completely fine-grained histories in the coördinate basis cannot be assigned probabilities; only suitable coarse-grained histories can. There are at least three common types of coarse graining: (1) specifying observables not at all times, but only at some times: (2)

specifying at any one time not a complete set of observables, but only some of them: (3) specifying for these observables not precise values, but only ranges of values. To illustrate all three, let us divide the $Q^i$ up into variables $x^i$ and $X^i$ and consider only sets of ranges $\{\Delta_\alpha^k\}$ of $x^i$ at times $t_k, k = 1, \cdots, n$. A set of alternatives at any one time consists of ranges $\Delta_\alpha^k$, which exhaust the possible values of $x^i$ as $\alpha$ ranges over all integers. An individual history is specified by particular $\Delta_\alpha$'s at the times $t_1, \cdots t_n$. We write $[\Delta_\alpha] = (\Delta_{\alpha_1}^1, \cdots \Delta_{\alpha_n}^n)$ for a particular history. A *set* of alternative histories is obtained by letting $\alpha_1 \cdots \alpha_n$ range over all values.

Let us use the same notation $[\Delta_\alpha]$ for the most general history that is a coarse graining of the completely fine-grained history in the coördinate basis, specified by ranges of the $Q^i$ at each time, including the possibility of full ranges at certain times, which eliminate those times from consideration.

(c)  Decohering Histories

The important theoretical construct for giving the rule that determines whether probabilities may be assigned to a given set of alternative histories, and what these probabilities are, is the decoherence functional $D\,[(\text{history})', (\text{history})]$. This is a complex functional on any pair of histories in the set. It is most transparently defined in the sum-over-histories framework for completely fine-grained history segments between an initial time $t_0$ and a final time $t_f$, as follows:

$$D\left[Q'^i(t), Q^i(t)\right] = \delta(Q_f'^i - Q_f^i) \exp\left\{i\left(S[Q'^i(t)] - S[Q^i(t)]\right)/\hbar\right\} \rho(Q_0'^i, Q_0^i) \quad . \tag{6}$$

Here $\rho$ is the initial density matrix of the universe in the $Q^i$ representation, $Q_0'^i$ and $Q_0^i$ are the initial values of the complete set of variables, and $Q_f'^i$ and $Q_f^i$ are the final values. The decoherence functional for coarse-grained histories is obtained from (6) according to the principle of superposition by summing over all that is not specified by the coarse graining. Thus,

$$D\left([\Delta_{\alpha'}], [\Delta_\alpha]\right) = \int\limits_{[\Delta_{\alpha'}]} \delta Q' \int\limits_{[\Delta_\alpha]} \delta Q \; \delta(Q_f'^i - Q_f^i) \; e^{i\left\{(S[Q'^i] - S[Q^i])/\hbar\right\}} \rho(Q_0'^i, Q_0^i) \quad . \tag{7}$$

More precisely, the integral is as follows (Figure 3): It is over all histories $Q'^i(t)$, $Q^i(t)$ that begin at $Q_0'^i, Q_0^i$ respectively, pass through the ranges $[\Delta_{\alpha'}]$ and $[\Delta_\alpha]$ respectively, and wind up at a common point $Q_f^i$ at any time $t_f > t_n$. It is completed by integrating over $Q_0'^i, Q_0^i$, and $Q_f^i$.

The connection between coarse-grained histories and completely fine-grained ones is transparent in the sum-over-histories formulation of quantum mechanics. However, the sum-over-histories formulation does not allow us to consider directly histories of the most general type. For the most general histories one needs to exploit directly the transformation theory of quantum mechanics and for this the Heisenberg picture is convenient. In the Heisenberg picture $D$ can be written

$$D\left([P_{\alpha'}], [P_\alpha]\right) = Tr\left[P_{\alpha_n'}^n(t_n) \cdots P_{\alpha_1'}^1(t_1) \rho P_{\alpha_1}^1(t_1) \cdots P_{\alpha_n}^n(t_n)\right] \quad . \tag{8}$$

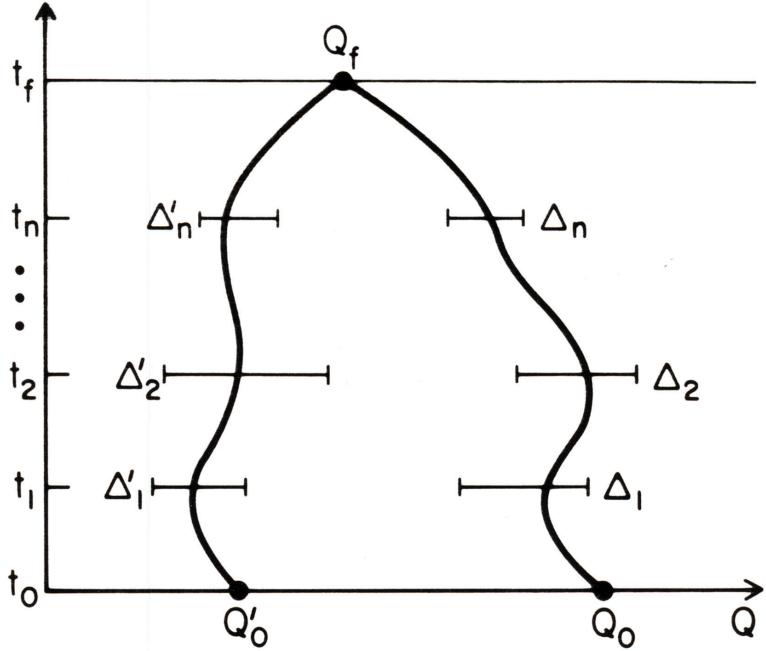

Figure 3. The sum-over-histories construction of the decoherence functional.

The projections in (8) are time ordered with the earliest on the inside. When the $P$'s are projections onto ranges $\Delta_\alpha^k$ of values of the $Q$'s, expressions (7) and (8) agree. From the cyclic property of the trace it follows that $D$ is always diagonal in the final indices $\alpha_n$ and $\alpha'_n$. (We assume throughout that the $P$'s are bounded operators in Hilbert space dealing, for example, with projections onto ranges of the $Q$'s and not onto definite values of the $Q$'s). Decoherence is thus an interesting notion only for strings of $P$'s that involve more than one time. Decoherence is automatic for "histories" that consist of alternatives at but one time.

Progressive coarse graining may be seen in the sum-over-histories picture as summing over those parts of the fine-grained histories not specified in the coarse-grained one, according to the principle of superposition. In the Heisenberg picture, eq. (8), the three common forms of coarse graining discussed above can be represented as follows: Summing on both sides of $D$ over all $P$'s at a given time and using (4) eliminates those $P$'s completely. Summing over all possibilities for certain variables at one time amounts to factoring the $P$'s and eliminating one of the factors by summing over it. Summing over ranges of values of a given variable at a given time corresponds to replacing the $P$'s for the partial ranges by one for the total range. Thus, if $[\overline{P_\beta}]$ is a coarse graining of the set of histories $\{[P_\alpha]\}$, we write

$$D\left([\overline{P_{\beta'}}], [\overline{P_\beta}]\right) = \sum_{\substack{\text{all } P'_\alpha \\ \text{not fixed by } [\overline{P_{\beta'}}]}} \quad \sum_{\substack{\text{all } P_\alpha \\ \text{not fixed by } [\overline{P_\beta}]}} D\left([P_{\alpha'}], [P_\alpha]\right) \quad . \tag{9}$$

In the most general case, we may think of the completely fine-grained limit as obtained from the coördinate representation by arbitrary unitary transformations at all times. All histories can be obtained by coarse-graining the various completely fine-grained ones, and coarse graining in its most general form involves taking arbitrary sums of $P$'s, as discussed earlier. We may use (9) in the most general case where $[\overline{P_\beta}]$ is a coarse graining of $[P_\alpha]$.

A set of coarse-grained alternative histories is said to *decohere* when the off-diagonal elements of $D$ are sufficiently small:

$$D\left([P_{\alpha'}],[P_\alpha]\right) \approx 0 \quad , \quad \text{for any } \alpha'_k \neq \alpha_k \quad . \tag{10}$$

This is a generalization of the condition for the absence of interference in the two-slit experiment (approximate equality of the two sides of (2)). It is a sufficient (although not a necessary) condition for the validity of the purely diagonal formula

$$D\left([\overline{P_\beta}],[\overline{P_\beta}]\right) \approx \sum_{\substack{\text{all } P_\alpha \text{ not} \\ \text{fixed by } [\overline{P_\beta}]}} D\left([P_\alpha],[P_\alpha]\right) \quad . \tag{11}$$

The rule for when probabilities can be assigned to histories of the universe is then this: To the extent that a *set* of alternative histories decoheres , probabilities can be assigned to its individual members. The probabilities are the *diagonal* elements of $D$. Thus,

$$\begin{aligned} p([P_\alpha]) &= D([P_\alpha],[P_\alpha]) \\ &= Tr\left[P^n_{\alpha_n}(t_n)\cdots P^1_{\alpha_1}(t_1)\rho P^1_{\alpha_1}(t_1)\cdots P^n_{\alpha_n}(t_n)\right] \end{aligned} \tag{12}$$

when the set decoheres. We will frequently write $p(\alpha_n t_n,\cdots\alpha_1 t_1)$ for these probabilities, suppressing the labels of the sets.

The probabilities defined by (11) obey the rules of probability theory as a consequence of decoherence. The principal requirement is that the probabilities be additive on "disjoint sets of the sample space". For histories this gives the sum rule

$$p\left([\overline{P_\beta}]\right) \approx \sum_{\substack{\text{all } P_\alpha \text{ not} \\ \text{fixed by } [\overline{P_\beta}]}} p\left([P_\alpha]\right) \tag{13}$$

These relate the probabilities for a set of histories to the probabilities for *all* coarser grained sets that can be constructed from it. For example, the sum rule eliminating all projections at only one time is

$$\begin{aligned} \sum_{\alpha_k} p(\alpha_n t_n,&\cdots\alpha_{k+1}t_{k+1},\alpha_k t_k,\alpha_{k-1}t_{k-1},\cdots,\alpha_1 t_1) \\ &\approx p(\alpha_n t_n,\cdots\alpha_{k+1}t_{k+1},\alpha_{k-1}t_{k-1},\cdots,\alpha_1 t_1) \quad . \end{aligned} \tag{14}$$

These rules follow trivially from (11) and (12). The other requirements from probability theory are that the probability of the whole sample space be unity, an easy consequence of (11) when complete coarse graining is performed, and that the probability for an empty set be zero, which means simply that the probability of any sequence containing a projection $P = 0$ must vanish, as it does.

The $p([P_\alpha])$ are *approximate* probabilities for histories, in the sense of Section II, up to the standard set by decoherence. Conversely, if a given standard for the probabilities is required by their use, it can be met by coarse graining until (10) and (13) are satisfied at the requisite level.

Further coarse graining of a decoherent set of alternative histories produces another set of decoherent histories since the probability sum rules continue to be obeyed. That is illustrated in Fig. 2, which makes it clear that in a progression from the trivial completely coarse graining to a completely fine graining, there are sets of histories where further fine graining always results in loss of decoherence. These are the *maximal* sets of alternative decohering histories.

These rules for probability exhibit another important feature: The operators in (12) are time-ordered. Were they not time-ordered (zig-zags) we could have assigned non-zero probabilities to conflicting alternatives at the same time. The time ordering thus expresses causality in quantum mechanics, a notion that is appropriate here because of the approximation of fixed background spacetime. The time ordering is related as well to the "arrow of time" in quantum mechanics, which we discuss below.

Given this discussion, the *fundamental formula* of quantum mechanics may be reasonably taken to be

$$D\left([P_{\alpha'}], [P_\alpha]\right) \approx \delta_{\alpha'_1 \alpha_1} \cdots \delta_{\alpha'_n \alpha_n} p([P_\alpha]) \tag{15}$$

for all $[P_\alpha]$ in a set of alternative histories. Vanishing of the off-diagonal elements of $D$ gives the rule for when probabilities may be consistently assigned. The diagonal elements give their values.

We could have used a weaker condition than (10) as the definition of decoherence, namely the necessary condition for the validity of the sum rules (11) of probability theory:

$$D\left([P_\alpha], [P_{\alpha'}]\right) + D\left([P_{\alpha'}], [P_\alpha]\right) \approx 0 \tag{16}$$

for any $\alpha'_k \neq \alpha_k$, or equivalently

$$Re\left\{D\left([P_\alpha], [P_{\alpha'}]\right)\right\} \approx 0 \quad . \tag{17}$$

This is the condition used by Griffiths[21] as the requirement for "consistent histories". However, while, as we shall see, it is easy to identify physical situations in which the off-diagonal elements of $D$ approximately vanish as the result of coarse graining, it is hard to think of a general mechanism that suppresses only their real parts. In the usual analysis of measurement, the off-diagonal parts of $D$ approximately vanish. We shall, therefore, explore the stronger condition (10) in what follows. That difference should not obscure the fact that in this part of our work we have reproduced what is essentially the approach of Griffiths[21], extended by Omnès[22].

(d)   Prediction and Retrodiction

Decoherent sets of histories are what we may discuss in quantum mechanics, for they may be assigned probabilities. Decoherence thus generalizes and replaces the notion of "measurement", which served this role in the Copenhagen interpretations. Decoherence is a more precise, more objective, more observer-independent idea. For example, if their associated histories decohere, we may assign probabilities to various values of reasonable scale density fluctuations in the early universe whether or not anything like a "measurement" was carried out on them and certainly whether or not there was an "observer" to do it. We shall return to a specific discussion of typical measurement situations in Section XI.

The joint probabilities $p(\alpha_n t_n, \cdots, \alpha_1 t_1)$ for the individual histories in a decohering set are the raw material for prediction and retrodiction in quantum cosmology. From them, the relevant conditional probabilities may be computed. The conditional probability of, one subset $\{\alpha_i t_i\}$, given the rest $\overline{\{\alpha_i t_i\}}$, is

$$p\left(\{\alpha_i t_i\}|\overline{\{\alpha_i t_i\}}\right) = \frac{p(\alpha_n t_n, \cdots, \alpha_1 t_1)}{p\left(\overline{\{\alpha_i t_i\}}\right)} \quad . \tag{18}$$

For example, the probability for *predicting* alternatives $\alpha_{k+1}, \cdots \alpha_n$, given that the alternatives $\alpha_1 \cdots \alpha_k$ have already happened, is

$$p(\alpha_n t_n, \cdots \alpha_{k+1} t_{k+1}|\alpha_k t_k, \cdots, \alpha_1 t_1) = \frac{p(\alpha_n t_n, \cdots, \alpha_1 t_1)}{p(\alpha_k t_k, \cdots \alpha_1 t_1)} \quad . \tag{19}$$

The probability that $\alpha_{n-1}, \cdots \alpha_1$ happened in the *past*, given present data summarized by an alternative $\alpha_n$ at the present time $t_n$, is

$$p(\alpha_{n-1} t_{n-1}, \cdots \alpha_1 t_1|\alpha_n t_n) = \frac{p(\alpha_n t_n, \cdots, \alpha_1 t_1)}{p(\alpha_n t_n)} \quad . \tag{20}$$

Decoherence ensures that the probabilities defined by (18) – (20) will approximately add to unity when summed over all remaining alternatives, because of (14).

Despite the similarity between (19) and (20), there are differences between prediction and retrodiction. Future predictions can all be obtained from an effective density matrix summarizing information about what has happened. If $\rho_{\text{eff}}$ is defined by

$$\rho_{\text{eff}} = \frac{P_{\alpha_k}^k(t_k) \cdots P_{\alpha_1}^1(t_1) \rho P_{\alpha_1}^1(t_1) \cdots P_{\alpha_k}^k(t_k)}{Tr[P_{\alpha_k}^k(t_k) \cdots P_{\alpha_1}^1(t_1) \rho P_{\alpha_1}^1(t_1) \cdots P_{\alpha_k}^k(t_k)]} \quad , \tag{21}$$

then

$$\begin{aligned} p(\alpha_n t_n, &\cdots \alpha_{k+1} t_{k+1}|\alpha_k t_k, \cdots, \alpha_1 t_1) \\ &= Tr[P_{\alpha_n}^n(t_n) \cdots P_{\alpha_{k+1}}^{k+1}(t_{k+1}) \rho_{\text{eff}} P_{\alpha_{k+1}}^{k+1}(t_{k+1}) \cdots P_{\alpha_n}^n(t_n)] \quad . \end{aligned} \tag{22}$$

By contrast, there is no effective density matrix representing present information from which probabilities for the past can be derived. As (20) shows, history requires knowledge of both present data *and* the initial condition of the universe.

Prediction and retrodiction differ in another way. Because of the cyclic property of the trace in (8), *any* final alternative decoheres and a probability can be predicted for it. By contrast we expect only certain variables to decohere in the past, appropriate to present data and the initial $\rho$. As the alternative histories of the electron in the two-slit experiment illustrate, there are many kinds of alternatives in the past for which the assignment of probabilities is prohibited in quantum mechanics. For those sets of alternatives that do decohere, the decoherence and the assigned probabilities typically will be approximate in the sense of Section II. It is unlikely, for example, that the initial state of the universe is such that the interference is exactly zero between two past positions of the sun in the sky.

These differences between prediction and retrodiction are aspects of the arrow of time in quantum mechanics. Mathematically they are consequences of the time ordering in (8) or (12). This time ordering does not mean that quantum mechanics singles out an absolute direction in time. Field theory is invariant under CPT. Performing a CPT transformation on (8) or (12) results in an equivalent expression in which the CPT-transformed $\rho$ is assigned to the far future and the CPT-transformed projections are anti-time-ordered. Either time ordering can, therefore, be used[c]; the important point is that there is a knowable Heisenberg $\rho$ from which probabilities can be predicted. It is by convention that we think of it as an "initial condition", with the projections in increasing time order from the inside out in (8) and (12).

While the formalism of quantum mechanics allows the universe to be discussed with either time ordering, the physics of the universe is time asymmetric, with a simple condition in what we call "the past." For example, the indicated present homogeneity of the thermodynamic arrow of time can be traced to the near homogeneity of the "early" universe implied by $\rho$ and the implication that the progenitors of approximately isolated subsystems started out far from equilibrium at "early" times.

Much has been made of the updating of the fundamental probability formula in (19) and in (21) and (22). By utilizing (21) the process of prediction may be organized so that for each time there is a $\rho_{\text{eff}}$ from which probabilities for the future may be calculated. The action of each projection, $P$, on both sides of $\rho$ in (21) along with the division by the appropriate normalizing factor is then sometimes called the "reduction of the wave packet". But this updating of probabilities is no different from the classical reassessment of probabilities that occurs after new information is obtained. In a sequence of horse races, the joint probability for the winners of eight races is converted, after the winners of the first three are known, into a reassessed probability for the remaining five races by exactly this process. The main thing is that, because of decoherence, the sum rules for probabilities are obeyed; once that is true, reassessment of probabilities is trivial.

The only non-trivial aspect of the situation is the choice of the string of $P$'s in (8) giving a decoherent set of histories.

(e)  Branches (Illustrated by a Pure $\rho$)

Decohering sets of alternative histories give a definite meaning to Everett's "branches". For a given such set of histories, the exhaustive set of $P_{\alpha_k}^k$ at each time $t_k$ corresponds to a branching.

To illustrate this even more explicitly, consider an initial density matrix that is a pure state, as in typical proposals for the wave function of the universe:

$$\rho = |\Psi><\Psi|   . \tag{23}$$

The initial state may be decomposed according to the projection operators that define the set of alternative histories

$$|\Psi> = \sum_{\alpha_1 \cdots \alpha_n} P_{\alpha_n}^n(t_n) \cdots P_{\alpha_1}^1(t_1)|\Psi>$$
$$\equiv \sum_{\alpha_1 \cdots \alpha_n} |[P_\alpha], \Psi>   . \tag{24}$$

The states $|[P_\alpha], \Psi>$ are approximately orthogonal as a consequence of their decoherence

$$<[P_{\alpha'}], \Psi|[P_\alpha], \Psi> \approx 0, \quad \text{for any} \quad \alpha_k' \neq \alpha_k   . \tag{25}$$

Eq. (25) is just a reëxpression of (10), given (23).

When the initial density matrix is pure, it is easily seen that some coarse graining in the present is always needed to achieve decoherence in the past. If the $P_{\alpha_n}^n(t_n)$ for the last time $t_n$ in (8) were all projections onto pure states, $D$ would factor for a pure $\rho$ and could never satisfy (10), except for certain special kinds of histories described near the end of Section VII, in which decoherence is automatic, independent of $\rho$. Similarly, it is not difficult to show that some coarse graining is required at any time in order to have decoherence of previous alternatives, with the same set of exceptions.

After normalization, the states $|[P_\alpha], \Psi >$ represent the individual histories or individual branches in the decohering set. We may, as for the effective density matrix of (d), summarize present information for prediction just by giving one of these states, with projections up to the present.

### (f) Sets of Histories with the Same Probabilities

If the projections $P$ are not restricted to a particular class (such as projections onto ranges of $Q^i$ variables), so that coarse-grained histories consist of arbitrary exhaustive families of projections operators, then the problem of exhibiting the decohering sets of strings of projections arising from a given $\rho$ is a purely algebraic one. Assume, for example, that the initial condition is known to be a pure state as in (23). The problem of finding ordered strings of exhaustive sets of projections $[P_\alpha]$ so that the histories $P_{\alpha_n}^n \cdots P_{\alpha_1}^1 |\Psi >$ decohere according to (25) is purely algebraic and involves just subspaces of Hilbert space. The problem is the same for one vector $|\Psi >$ as for any other. Indeed, using subspaces that are *exactly* orthogonal, we may identify sequences that *exactly* decohere.

However, it is clear that the solution of the mathematical problem of enumerating the sets of decohering histories of a given Hilbert space has no physical content by itself. No description of the histories has been given. No reference has been made to a theory of the fundamental interactions. No distinction has been made between one vector in Hilbert space as a theory of the initial condition and any other. The resulting probabilities, which can be calculated, are merely abstract numbers.

We obtain a description of the sets of alternative histories of the universe when the operators corresponding to the fundamental fields are identified. We make contact with the theory of the fundamental interactions if the evolution of these fields is given by a fundamental Hamiltonian. Different initial vectors in Hilbert space will then give rise to decohering sets having different descriptions in terms of the fundamental fields. The probabilities acquire physical meaning.

Two different simple operations allow us to construct from one set of histories another set with a *different description* but the *same probabilities*. First consider unitary transformations of the $P$'s that are constant in time and leave the initial $\rho$ fixed

$$\rho = U\rho U^{-1} \quad , \tag{26}$$

$$\tilde{P}_\alpha^k(t) = U P_\alpha^k(t) U^{-1} \quad . \tag{27}$$

If $\rho$ is pure there will be very many such transformations; the Hilbert space is large and only a single vector is fixed. The sets of histories made up from the $\{\tilde{P}_\alpha^k\}$ will have an identical decoherence functional to the sets constructed from the corresponding $\{P_\alpha^k\}$. If one set decoheres, the other will and the probabilities for the individual histories will be the same.

In a similar way, decoherence and probabilities are invariant under arbitrary reassignments of the times in a string of $P$'s (as long as they continue to be ordered), with the projection operators at the altered times unchanged as operators in Hilbert space. This is because in the Heisenberg picture every projection is at *any* time a projection operator for *some* quantity.

The histories arising from constant unitary transformations or from reassignment of times of a given set of $P$'s will, in general, have very different descriptions in terms of fundamental fields from that of the original set. We are considering transformations such as (27) in an active (or alibi) sense so that the field operators and Hamiltonian are unchanged. (The passive (or alias) transformations, in which these are transformed, are easily understood.)    A set of projections onto the ranges of field values in a spatial region is generally transformed by (27) or by any reassignment of the times into an extraordinarily complicated combination of all fields and all momenta at all positions in the universe! Histories consisting of projections onto values of similar quantities at different times can thus become histories of very different quantities at various other times.

In ordinary presentations of quantum mechanics, two histories with different descriptions can correspond to physically distinct situations because it is presumed that various different Hermitian combinations of field operators are potentially measurable by different kinds of external apparatus. In quantum cosmology, however, apparatus and system are considered together and the notion of physically distinct situations may have a different character.

## V. THE ORIGINS OF DECOHERENCE

What are the features of coarse-grained sets of histories that decohere, given the $\rho$ and $H$ of the universe? In seeking to answer this question it is important to keep in mind the basic aspects of the theoretical framework on which decoherence depends. Decoherence of a set of alternative histories is not a property of their operators *alone*. It depends on the relations of those operators to the density matrix $\rho$. Given $\rho$, we could, in principle, *compute* which sets of alternative histories decohere.

We are not likely to carry out a computation of all decohering sets of alternative histories for the universe, described in terms of the fundamental fields, anytime in the near future, if ever. However, if we focus attention on coarse grainings of particular variables, we can exhibit widely occurring mechanisms by which they decohere in the presence of the actual $\rho$ of the universe. We have mentioned in Section IVc that decoherence is automatic if the projection operators $P$ refer only to one time; the same would be true even for different times if all the $P$'s commuted with one another. Of course, in cases of interest, each $P$ typically factors into commuting projection operators, and the factors of $P$'s for different times often fail to commute with one another, for example factors that are projections onto related ranges of values of the same Heisenberg operator at different times. However, these non-commuting factors may be correlated, given $\rho$, with other projection factors that do commute or, at least, effectively commute inside the trace with the density matrix $\rho$ in Equation (8) for the decoherence functional. In fact, these other projection factors may commute with all the subsequent $P$'s and thus allow themselves to be moved to the outside of the trace formula. When all the non-commuting factors are correlated in this manner with effectively commuting ones, then the off-diagonal terms in the decoherence functional vanish, in other words, decoherence results. Of course, all this behavior may be approximate, resulting in approximate decoherence.

This type of situation is fundamental in the interpretation of quantum mechanics. Non-commuting quantities, say at different times, may be correlated with commuting or effectively commuting quantities because of the character of $\rho$ and $H$, and thus produce decoherence of strings of $P$'s despite their non-commutation. For a pure $\rho$, for example, the behavior of the effectively commuting variables leads to the orthogonality of the branches of the state $|\Psi>$, as defined in (24). We shall see that correlations of this character are central to understanding historical records (Section X) and measurement situations (Section XI).

As an example of decoherence produced by this mechanism, consider a coarse-grained set of histories defined by time sequences of alternative approximate localizations of a massive body such as a planet or even a typical interstellar dust grain. As shown by Joos and Zeh,[20] even if the successive localizations are spaced as closely as a nanosecond, such histories decohere as a consequence of scattering by the $3°$ cosmic background radiation (if for no other reason). Different positions become correlated with nearly orthogonal states of the photons. More importantly, each alternative sequence of positions becomes correlated with a different orthogonal state of the photons at the final time. This accomplishes the decoherence and we may loosely say that such histories of the position of a massive body are "decohered" by interaction with the photons of the background radiation.

Other specific models of decoherence have been discussed by many authors, among them Joos and Zeh,[20] Caldeira and Leggett,[25] and Žurek.[26] Typically these discussions have focussed on a coarse graining that involves only certain variables analogous to the position variables above. Thus the emphasis is on particular non-commuting factors of the projection operators and not on correlated operators that may be accomplishing the approximate decoherence. Such coarse grainings do not, in general, yield the most refined approximately decohering sets of histories, since one could include projections onto ranges of values of the correlated operators without losing the decoherence.

The simplest model consists of a single oscillator interacting bilinearly with a large number of others, and a coarse graining which involves only the coördinates of the special oscillator. Let $x$ be the coördinate of the special oscillator, $M$ its mass, $\omega_R$ its frequency renormalized by its interactions with the others, and $S_{\text{free}}$ its free action. Consider the special case where the density matrix of the whole system, referred to an initial time, factors into the product of a density matrix $\bar{\rho}(x', x)$ of the distinguished oscillator and another for the rest. Then, generalizing slightly a treatment of Feynman and Vernon[27], we can write $D$ defined by (7) as

$$D\left([\Delta_{\alpha'}], [\Delta_\alpha]\right) = \int_{[\Delta_{\alpha'}]} \delta x'(t) \int_{[\Delta_\alpha]} \delta x(t) \delta(x'_f - x_f)$$

$$\exp\left\{ i\Big(S_{\text{free}}[x'(t)] - S_{\text{free}}[x(t)] + W[x'(t), x(t)]\Big)/\hbar \right\} \bar{\rho}(x'_0, x_0) \,, \qquad (28)$$

the intervals $[\Delta_\alpha]$ referring here only to the variables of the distinguished oscillator. The sum over the rest of the oscillators has been carried out and is summarized by the Feynman - Vernon influence functional $\exp(iW[x'(t), x(t)])$. The remaining sum over $x'(t)$ and $x(t)$ is as in (7).

The case when the other oscillators are in an initial thermal distribution has been extensively investigated by Caldeira and Leggett.[25] In the simple limit of a uniform continuum of oscillators cut off at frequency $\Omega$ and in the Fokker - Planck limit of $kT \gg \hbar\Omega \gg \hbar\omega_R$,

they find

$$W[x'(t), x(t)] = -M\gamma \int dt [x'\dot{x}' - x\dot{x} + x'\dot{x} - x\dot{x}']$$
$$+ i\frac{2M\gamma kT}{\hbar} \int dt [x'(t) - x(t)]^2 \quad, \tag{29}$$

where $\gamma$ summarizes the interaction strengths of the distinguished oscillator with its environment. The real part of $W$ contributes dissipation to the equations of motion. The imaginary part squeezes the trajectories $x(t)$ and $x'(t)$ together, thereby providing approximate decoherence. Very roughly, primed and unprimed position intervals differing by distances $d$ on opposite sides of the trace in (28) will decohere when spaced in time by intervals

$$t \gtrsim \frac{1}{\gamma} \left[ \left( \frac{\hbar}{\sqrt{2MkT}} \right) \cdot \left( \frac{1}{d} \right) \right]^2 \quad. \tag{30}$$

As stressed by Żurek[26], for typical macroscopic parameters this minimum time for decoherence can be many orders of magnitude smaller than a characteristic dynamical time, say the damping time $1/\gamma$. (The ratio is around $10^{-40}$ for $M \sim$ gm, $T \sim 300°K, d \sim$ cm!)

The behavior of a coarse-grained set of alternative histories based on projections, at times spaced far enough apart for decoherence, onto ranges of values of $x$ alone, is then roughly classical in that the successive ranges of positions follow roughly classical orbits, but with the pattern of classical correlation disturbed by various effects, especially (a) the effect of quantum spreading of the $x$-coördinate, (b) the effect of quantum fluctuations of the other oscillators, and (c) classical statistical fluctuations, which are lumped with (b) when we use the fundamental formula. We see that the larger the mass $M$, the shorter the decoherence time and the more the $x$-coördinate resists the various challenges to its classical behavior.

What the above models convincingly show is that decoherence will be widespread in the universe for certain familiar "classical" variables. The answer to Fermi's question to one of us of why we don't see Mars spread out in a quantum superposition of different positions in its orbit is that such a superposition would rapidly decohere. We now proceed to a more detailed discussion of such decoherence.

## VI. QUASICLASSICAL DOMAINS

As observers of the universe, we deal with coarse grainings that are appropriate to our limited sensory perceptions, extended by instruments, communication, and records, but in the end characterized by a great amount of ignorance. Yet we have the impression that the universe exhibits a finer-grained set of decohering histories, independent of us, defining a sort of "classical domain", governed largely by classical laws, to which our senses are adapted while dealing with only a small part of it. No such coarse graining is determined by pure quantum theory alone. Rather, like decoherence, the existence of a quasiclassical domain in the universe must be a consequence of its initial condition and the Hamiltonian describing its evolution.

Roughly speaking, a quasiclassical domain should be a set of alternative decohering histories, maximally refined consistent with decoherence, with its individual histories exhibiting as much as possible patterns of classical correlation in time. Such histories cannot be *exactly* correlated in time according to classical laws because sometimes their classical evolution is disturbed by quantum events. There are no classical domains, only quasiclassical ones.

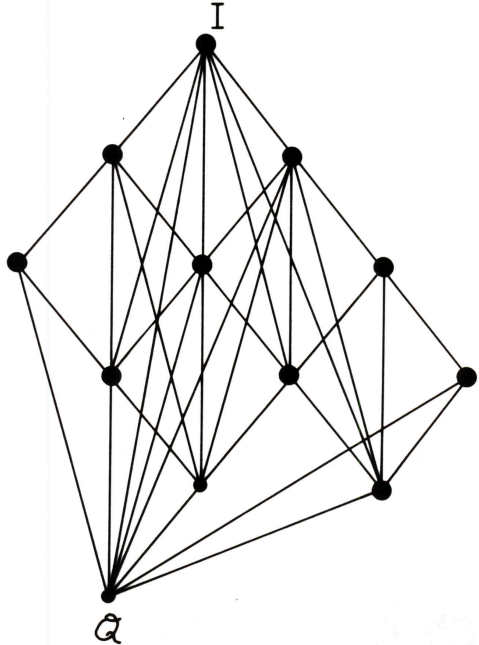

Figure 4. If the completely fine grained histories arise from a single complete set of observables, say the set $\mathcal{Q}$ of field variables $Q^i$ at each point in space and every time, then the possible coarse-grained histories will be a subset of those illustrated in Figure 2. Maximal sets can still be defined but will, in general, differ from those of Figure 2.

We wish to make the question of the existence of one or more quasiclassical domains into a *calculable* question in quantum cosmology and for this we need criteria to measure how close a set of histories comes to constituting a "classical domain". We have not solved this problem to our satisfaction, but, in the next few Sections, we discuss some ideas that may contribute toward its solution.

## VII. MAXIMAL SETS OF DECOHERING HISTORIES

Decoherence results from coarse graining. As described in Section IVb and Fig. 2, coarse grainings can be put into a partial ordering with one another. A set of alternative histories is a coarse graining of a finer set if all the exhaustive sets of projections $\{P_\alpha^k\}$ making up the coarser set of histories are obtained by partial sums over the projections making up the finer set of histories.

Maximal sets of alternative decohering histories are those for which there are no finer-grained sets that are decoherent. It is desirable to work with maximal sets of decohering alternative histories because they are not limited by the sensory capacity of any set of observers – they can cover phenomena in all parts of the universe and at all epochs that could be observed, whether or not any observer was present. Maximal sets are the most refined descriptions of the universe that may be assigned probabilities in quantum mechanics.

The class of maximal sets possible for the universe depends, of course, on the completely fine-grained histories that are presented by the actual quantum theory of the universe. If we utilize to the full, at each moment of time, all the projections permitted by transformation theory, which gives quantum mechanics its protean character, then there is an infinite variety of completely fine-grained sets, as illustrated in Figure 2. However, were there some fundamental reason to restrict the completely fine grained sets, as would be the case if sum-over-histories quantum mechanics were fundamental, then the class of maximal sets would be smaller as illustrated in Figure 4. We shall proceed as if all fine grainings are allowed.

If a full correlation exists between a projection in a coarse graining and another projection not included, then the finer graining including both still defines a decoherent set of histories. In a maximal set of decoherent histories, both correlated projections must be included if either one is included. Thus, in the mechanism of decoherence discussed in Section V, projections onto the correlated orthogonal states of the $3°$K photons are included in the maximal set of decohering histories along with the positions of the massive bodies. Any projections defining historical records such as we shall describe in Section X, or values of measured quantities such as we shall describe in Section XI, must similarly be included in a maximal set.

More information about the initial $\rho$ and $H$ is contained in the probabilities of a finer-grained set of histories than in those of a coarser grained set. It would be desirable to have a quantitative measure of *how much* more information is obtained in a further fine graining of a coarse-grained set of alternative histories. Such a quantity would then measure how much closer a decoherent fine graining comes to maximality in a physically relevant sense.

We shall discuss a quantity that, while not really a measure of maximality, is useful in exploring some aspects of it. In order to construct that quantity the usual entropy formula is applied to sets of alternative decohering *histories* of the universe, rather than, as more usually, alternatives at a single time. We make use of the coarse-grained density matrix $\tilde{\rho}$ defined using the methods of Jaynes[28], *but generalized to take account of the density matrix of the universe and applied to the probabilities for histories*. The density matrix $\tilde{\rho}$ is constructed by maximizing the entropy functional

$$S(\tilde{\rho}) = -Tr(\tilde{\rho} \log \tilde{\rho}) \tag{31}$$

over all density matrices $\tilde{\rho}$ that satisfy the constraints ensuring that each

$$Tr\left[ P^n_{\alpha'_n}(t_n) \cdots P^1_{\alpha'_1}(t_1) \tilde{\rho} P^1_{\alpha_1}(t_1) \cdots P^n_{\alpha_n}(t_n) \right] \tag{32}$$

has the same value it would have had when computed with the density matrix of the universe, $\rho$, for a given set of coarse-grained histories. The density matrix $\tilde{\rho}$ thus reproduces the decoherence functional for this set of histories, and in particular their probabilities, but possesses as little information as possible beyond those properties.

A fine graining of a set of alternative histories leads to *more* conditions on $\tilde{\rho}$ of the form (32) than in the coarser-grained set. In non-trivial cases $S(\tilde{\rho})$ is, therefore, lowered and $\tilde{\rho}$ becomes closer to $\rho$.

If the insertion of apparently new $P$'s into a chain is redundant, then $S(\tilde{\rho})$ will not be lowered. A simple example will help to illustrate this: Consider the set of histories consisting of projections $P^m_{\alpha_m}(t_m)$ which project onto an orthonormal basis for Hilbert space at one time, $t_m$. Trivial further decoherent fine grainings can be constructed as follows: At each other time $t_k$ introduce a set of projections $P^k_{\alpha_k}(t_k)$ that, through the equations of motion, are identical operators in Hilbert space to the set $P^m_{\alpha_m}(t_m)$. In this way, even though we are going through the motions of introducing a completely fine-grained set of histories covering all the times, we are really just repeating the projections $P^m_{\alpha_m}(t_m)$ over and over again. We thus have a completely fine-grained set of histories that, in fact, consists of just one fine-grained set of projections and decoheres exactly because there is only one such set. Indeed, in terms of $S(\tilde{\rho})$ it is no closer to maximality than the set consisting of $P^m_{\alpha_m}(t_m)$ at one time. The quantity $S(\tilde{\rho})$ thus serves to identify such trivial refinements which amount to redundancy in the conditions (32).

We can generalize the example in an interesting way by constructing the special kinds of histories mentioned after Equation (25). We take $t_m$ to be the final time and then adjoin, at earlier and earlier times, a succession of progressive coarse grainings of the set $\{P^m_{\alpha_m}(t_m)\}$. Thus, as time moves forward, the only projections are finer and finer grainings terminating in the one-dimensional $P^m_{\alpha_m}(t_m)$. We thus have again a set of histories in which decoherence is automatic, independent of the character of $\rho$ and for which $S(\tilde{\rho})$ has the same value it would have had if only the conditions at the final time were considered.

In a certain sense, $S(\tilde{\rho})$ for histories can be regarded as decreasing with time. If we consider $S(\tilde{\rho})$ for a string of alternative projections up to a certain time $t_n$, as in (32), and then adjoin an additional set of projections for a later time, the number of conditions on $\tilde{\rho}$ is increased and thus the value of $S(\tilde{\rho})$ is decreased (or, in trivial cases, unchanged). That is natural, since $S(\tilde{\rho})$ is connected with the lack of information contained in a set of histories and that information increases with non-trivial fine graining of the histories, no matter what the times for which the new $P$'s are introduced. (In some related areas, a quantity like $S$ that keeps decreasing as a result of adjoining projections at later times can be converted into an increasing quantity by adding an algorithmic complexity term.[29])

The quantity $S(\tilde{\rho})$ is closely related to other fundamental quantities in physics. One can show, for example, that when *used with the $\rho_{\mathrm{eff}}$ representing present data and with alternatives at a single time*, these techniques give a unified and generalized treatment of the variety of coarse grainings commonly introduced in statistical mechanics; and, as Jaynes and others have pointed out, the resulting $S(\tilde{\rho})$'s are the physical entropies of statistical mechanics. Here, however, these techniques are applied to time *histories* and the initial condition is utilized. The quantity $S(\tilde{\rho})$ is also related to the notion of thermodynamic depth currently being investigated by Lloyd.[30]

## VIII. CLASSICITY

Some maximal sets will be more nearly classical than others. The more nearly classical sets of histories will contain projections (onto related ranges of values) of operators, for different times, that are connected to one another by the unitary transformations $e^{-iH(\Delta t)}$ and that are correlated for the most part along classical paths, with probabilities near zero and one for the successive projections. This pattern of classical correlation may be disturbed by the inclusion, in the maximal set of projection operators, of other variables, which do not behave in this way (as in measurement situations to be described later). The pattern may also be disturbed by quantum spreading and by quantum and classical fluctuations, as described in connection with the oscillator example treated in Section V. Thus we can, at best, deal with *quasiclassical* maximal sets of alternative decohering histories, with trajectories that split and fan out as a result of the processes that make the decoherence possible. As we stressed earlier, there are no classical domains, only quasiclassical ones.

The impression that there is something like a classical domain suggests that we try to define quasiclassical domains precisely by searching for a measure of classicity for each of the maximal sets of alternative decohering histories and concentrating on the one (or ones) with maximal classicity. Such a measure would be applied to the elements of $D$ and the corresponding coarse graining. It should favor predictability, involving patterns of classical correlation as described above. It should also favor maximal sets of alternative decohering histories that are relatively fine-grained as opposed to those which had to be carried to very coarse graining before they would give decoherence. We are searching for such a measure. It should provide a precise and quantitative meaning to the notion of quasiclassical domain.

## IX. QUASICLASSICAL OPERATORS

What are the projection operators that specify the coarse graining of a maximal set of alternative histories with high classicity, which defines a quasiclassical domain? They will include, as mentioned above, projections onto comparable ranges of values of certain operators at sequences of times, obeying roughly classical equations of motion, subject to fluctuations that cause their trajectories to fan out from time to time. We can refer to these operators, which habitually decohere, as "quasiclassical operators". What these quasi-classical operators are, and how many of them there are, depends not only on $H$ and $\rho$, but also on the epoch, on the spatial region, and on previous branchings.

We can understand the origin of at least some quasiclassical operators in reasonably general terms as follows: In the earliest instants of the universe the operators defining spacetime on scales well above the Planck scale emerge from the quantum fog as quasi-classical.[31] Any theory of the initial condition that does not imply this is simply inconsistent with observation in a manifest way. The background spacetime thus defined obeys the Einstein equation. Then, where there are suitable conditions of low temperature, etc., various sorts of hydrodynamic variables may emerge as quasi-classical operators. These are integrals over suitable small volumes of densities of conserved or nearly conserved quantities. Examples are densities of energy, momentum, baryon number, and, in later epochs, nuclei, and even chemical species. The sizes of the volumes are limited above by maximality and are limited below by classicity because they require sufficient "inertia" to enable them to resist deviations from predictability caused by their interactions with one another, by quantum spreading, and by the quantum and statistical fluctuations resulting from interactions with the rest of the universe. Suitable integrals of densities of approximately conserved quantities are thus candidates for habitually decohering quasi-classical operators. Field theory is local, and it is an interesting question whether that locality somehow picks out local densities as the source of habitually decohering quantities. It is hardly necessary to note that such hydrodynamic variables are among the principal variables of classical physics.[32]

In the case of densities of conserved quantities, the integrals would not change at all if the volumes were infinite. For smaller volumes we expect approximate persistence. When, as in hydrodynamics, the rates of change of the integrals form part of an approximately closed system of equations of motion, the resulting evolution is just as classical as in the case of persistence.

# X. BRANCH DEPENDENCE

As the discussion in Sections V and IX shows, physically interesting mechanisms for decoherence will operate differently in different alternative histories for the universe. For example, hydrodynamic variables defined by a relatively small set of volumes may decohere at certain locations in spacetime in those branches where a gravitationally condensed body (e.g. the earth) actually exists, and may not decohere in other branches where no such condensed body exists at that location. In the latter branch there simply may be not enough "inertia" for densities defined with too small volumes to resist deviations from predictability. Similarly, alternative spin directions associated with Stern-Gerlach beams may decohere for those branches on which a photographic plate detects their beams and not in a branch where they recombine coherently instead. There are no variables that are expected to decohere universally. Even the mechanisms causing spacetime geometry at a given location to decohere on scales far above the Planck length cannot necessarily be expected to operate in the same way on a branch where the location is the center of a black hole as on those branches where there is no black hole nearby.

How is such "branch dependence" described in the formalism we have elaborated? It is not described by considering histories where the *set* of alternatives at one time (the $k$ in a set of $P_\alpha^k$) depends on *specific* alternatives (the $\alpha$'s) of sets of earlier times. Such dependence would destroy the derivation of the probability sum rules from the fundamental formula. However, there is no such obstacle to the set of alternatives at one time depending on the *sets* of alternatives at all previous times. It is by exploiting this possibility, together with the possibility of present records of past events, that we can correctly describe the sense in which there is branch dependence of decoherence, as we shall now discuss.

A record is a present alternative that is, with high probability, correlated with an alternative in the past. The construction of the relevant probabilities was discussed in Section IV, including their dependence on the initial condition of the universe (or at least on information that effectively bears on that initial condition). The subject of history is most honestly described as the construction of probabilities for the past, given such records. Even non-commuting alternatives such as a position and its momentum at different, even nearby times may be stored in presently commuting record variables.

The branch dependence of histories becomes explicit when sets of alternatives are considered that include records of specific events in the past. To illustrate this, consider the example above, where different sorts of hydrodynamic variables might decohere or not depending on whether there was a gravitational condensation. The set of alternatives that decohere must refer both to the records of the condensation *and* to hydrodynamic variables. Hydrodynamic variables with smaller volumes would be part of the subset with the record that the condensation took place and vice versa.

The branch dependence of decoherence provides the most direct argument against the position that a classical domain should simply be *defined* in terms of a certain set of variables (e.g. values of spacetime averages of the fields in the classical action). There are unlikely to be any physically interesting variables that decohere independent of circumstance.

# XI. MEASUREMENT SITUATIONS

When a correlation exists between the ranges of values of two operators of a quasiclassical domain, there is a *measurement situation*. From a knowledge of the value of one, the value of the other can be deduced because they are correlated with probability near unity. Any such correlation exists in some branches of the universe and not in others; for example, measurements in a laboratory exist only in those branches where the laboratory was actually constructed!

We use the term "measurement situation" rather than "measurement" for such correlations to stress that nothing as sophisticated as an "observer" need be present for them to exist. If there are many significantly different quasiclassical domains, different measurement situations may be exhibited by each one.

When the correlation we are discussing is between the ranges of values of two quasiclassical operators that *habitually* decohere, as discussed in Section IX, we have a measurement situation of a familiar classical kind. However, besides the quasiclassical operators, the highly classical maximal sets of alternative histories of a quasiclassical domain may include *other* operators having ranges of values strongly correlated with the quasiclassical ones at particular times. Such operators, not normally decohering are, in fact, included among the decohering set only by virtue of their correlation with a habitually decohering one. In this case we have a measurement situation of the kind usually discussed in quantum mechanics. Suppose, for example, in the inevitable Stern-Gerlach experiment, that $\sigma_z$ of a spin-1/2 particle is correlated with the orbit of an atom in an inhomogeneous magnetic field. If the two orbits decohere because of interaction with something else (the atomic excitations in a photographic plate for example), then the spin direction will be included in the maximal set of decoherent histories, fully correlated with the decohering orbital directions. The spin direction is thus measured.

The recovery of the Copenhagen rule for when probabilities may be assigned is immediate. Measured quantities are correlated with decohering histories. Decohering histories can be assigned probabilities. Thus in the two-slit experiment (Figure 1), when the electron interacts with an apparatus that determines which slit it passed through, it is the decoherence of the alternative configurations of the apparatus that enables probabilities to be assigned for the electron.

Correlation between the ranges of values of operators of a quasiclassical domain is the *only* defining property of a measurement situation. Conventionally, measurements have been characterized in other ways. Essential features have been seen to be irreversibility, amplification beyond a certain level of signal-to-noise, association with a macroscopic variable, the possibility of further association with a long chain of such variables, and the formation of enduring records. Efforts have been made to attach some degree of precision to words like "irreversible", "macroscopic", and "record", and to discuss what level of "amplification" needs to be achieved.[33] While such characterizations of measurement are difficult to define precisely,[d] some can be seen in a rough way to be consequences of the definition that we are attempting to introduce here, as follows:

Correlation of a variable with the quasiclassical domain (actually, inclusion in its set of histories) accomplishes the amplification beyond noise and the association with a macroscopic variable that can be extended to an indefinitely long chain of such variables. The relative predictability of the classical world is a generalized form of record. The approximate constancy of, say, a mark in a notebook is just a special case; persistence in a classical orbit is just as good.

Irreversibility is more subtle. One measure of it is the cost (in energy, money, etc.) of tracking down the phases specifying coherence and restoring them. This is intuitively large in many typical measurement situations. Another, related measure is the negative of the logarithm of the probability of doing so. If the probability of restoring the phases in any particular measurement situation were significant, then we would not have the necessary amount of decoherence. The correlation could not be inside the set of decohering histories. Thus, this measure of irreversibility is large. Indeed, in many circumstances where the phases are carried off to infinity or lost in photons impossible to retrieve, the probability of recovering them is truly zero and the situation perfectly irreversible — infinitely costly to reverse and with zero probability for reversal!

Defining a measurement situation solely as the existence of correlations in a quasiclassical domain, if suitable general definitions of maximality and classicity can be found, would have the advantages of clarity, economy, and generality. Measurement situations occur throughout the universe and without the necessary intervention of anything as sophisticated as an "observer". Thus, by this definition, the production of fission tracks in mica deep in the earth by the decay of a uranium nucleus leads to a measurement situation in a quasiclassical domain in which the tracks directions decohere, whether or not these tracks are ever registered by an "observer".

## XII. COMPLEX ADAPTIVE SYSTEMS

Our picture is of a universe that, as a consequence of a particular initial condition and of the underlying Hamiltonian, exhibits at least one quasiclassical domain made up of suitably defined maximal sets of alternative histories with as much classicity as possible. The quasiclassical domains would then be a consequence of the theory and its boundary condition, not an artifact of our construction. How do we then characterize our place as a collectivity of observers in the universe?

Both singly and collectively we are examples of the general class of complex adaptive systems. When they are considered within quantum mechanics as portions of the universe, making observations, we refer to such complex adaptive systems as information gathering and utilizing systems ($IGUS$es). The general characterization of complex adaptive systems is the subject of much ongoing research, which we cannot discuss here. From a quantum-mechanical point of view the foremost characteristic of an $IGUS$ is that, in some form of approximation, however crude or classical, it employs the fundamental formula, with what amounts to a rudimentary theory of $\rho$, $H$, and quantum mechanics. Probabilities of interest to the $IGUS$ include those for correlations between its memory and the external world. (Typically these are assumed perfect; not always such a good approximation!) The approximate fundamental formula is used to compute probabilities on the basis of present data, make predictions, control future perceptions on the basis of these predictions (i.e., exhibit behavior), acquire further data, make further predictions, and so on.

To carry on in this way, an $IGUS$ uses probabilities for histories referring both to the future and the past. An $IGUS$ uses decohering sets of alternative histories and therefore performs further coarse graining on a quasiclassical domain. Naturally, its coarse graining is very much coarser than that of the quasiclassical domain since it utilizes only a few of the variables in the universe.

The reason such systems as $IGUS$es exist, functioning in such a fashion, is to be sought in their evolution within the universe. It seems likely that they evolved to make predictions because it is adaptive to do so.[e] The reason, therefore, for their focus on decohering variables is that these are the *only* variables for which predictions can be made. The reason for their focus on the histories of a quasiclassical domain is that these present enough regularity over time to permit the generation of models (schemata) with significant predictive power.

If there is essentially only one quasiclassical domain, then naturally the $IGUS$ utilizes further coarse grainings of it. If there are many essentially inequivalent quasiclassical domains, then we could adopt a subjective point of view, as in some traditional discussions of quantum mechanics, and say that the $IGUS$ "chooses" its coarse graining of histories and, therefore, "chooses" a particular quasiclassical domain, or a subset of such domains, for further coarse graining. It would be better, however, to say that the $IGUS$ evolves to exploit a particular quasiclassical domain or set of such domains. Then $IGUS$es, including human beings, occupy no special place and play no preferred role in the laws of physics. They merely utilize the probabilities presented by quantum mechanics in the context of a quasiclassical domain.

# XIII. CONCLUSIONS

We have sketched a program for understanding the quantum mechanics of the universe and the quantum mechanics of the laboratory, in which the notion of quasiclassical domain plays a central role. To carry out that program, it is important to complete the definition of a quasiclassical domain by finding the general definition for classicity. Once that is accomplished, the question of how many and what kinds of essentially inequivalent quasiclassical domains follow from $\rho$ and $H$ becomes a topic for serious theoretical research. So is the question of what are the general properties of $IGUS$es that can exist in the universe exploiting various quasiclassical domains, or the unique one if there is essentially only one.

It would be a striking and deeply important fact of the universe if, among its maximal sets of decohering histories, there were one roughly equivalent group with much higher classicities than all the others. That would then be *the* quasiclassical domain, completely independent of any subjective criterion, and realized within quantum mechanics by utilizing only the initial condition of the universe and the Hamiltonian of the elementary particles.

Whether the universe exhibits one or many maximal sets of branching alternative histories with high classicities, those quasiclassical domains are the possible arenas of prediction in quantum mechanics.

It might seem at first sight that in such a picture the complementarity of quantum mechanics would be lost; in a given situation, for example, *either* a momentum *or* a coördinate could be measured, leading to different kinds of histories. We believe that impression is illusory. The histories in which an observer, as part of the universe, measures $p$ and the histories in which that observer measures $x$ are decohering alternatives. The important point is that the decoherent histories of a quasiclassical domain contain all possible choices that might be made by all possible observers that might exist, now, in the past, or in the future for that domain.

The EPR or EPRB situation is no more mysterious. There, a choice of measurements, say, $\sigma_x$ or $\sigma_y$ for a given electron, is correlated with the behavior of $\sigma_x$ or $\sigma_y$ for another electron because the two together are in a singlet spin state even though widely separated. Again, the two measurement situations (for $\sigma_x$ and $\sigma_y$) decohere from each other. but here, in each, there is also a correlation between the information obtained about one spin and the information that can be obtained about the other. This behavior, although unfortunately called "non-local" by some authors, involves no non-locality in the ordinary sense of quantum field theory and no possibility of signaling outside the light cone. The problem with the "local realism" that Einstein would have liked is not the locality but the *realism*. Quantum mechanics describes *alternative* decohering histories and one cannot assign "reality" simultaneously to different alternatives because they are contradictory. Everett[9] and others[11] have described this situation, not incorrectly, but in a way that has confused some, by saying that the histories are all "equally real" (meaning only that quantum mechanics prefers none over another except via probabilities) and by referring to "many worlds" instead of "many histories".

We conclude that resolution of the problems of interpretation presented by quantum mechanics is not to be accomplished by further intense scrutiny of the subject as it applies to reproducible laboratory situations, but rather through an examination of the origin of the universe and its subsequent history. Quantum mechanics is best and most fundamentally understood in the context of quantum cosmology. The founders of quantum mechanics were right in pointing out that something external to the framework of wave function and Schrödinger equation *is* needed to interpret the theory. But it is not a postulated classical world to which quantum mechanics does not apply. Rather it is the initial condition of the universe that, together with the action function of the elementary particles and the throws of quantum dice since the beginning, explains the origin of quasiclassical domain(s) within quantum theory itself.

# ACKNOWLEDGMENTS

One of us, MG-M, would like to acknowledge the great value of conversations about the meaning of quantum mechanics with Felix Villars and Richard Feynman in 1963-64 and again with Richard Feynman in 1987-88. He is also very grateful to Valentine Telegdi for discussions during 1985-86, which persuaded him to take up the subject again after twenty years. Both of us are indebted to Telegdi for further interesting conversations since 1987. We would also like to thank R. Griffiths for a useful communication and a critical reading of the manuscript and R. Penrose for a helpful discussion.

Part of this work was carried out at various times at the Institute for Theoretical Physics, Santa Barbara, the Aspen Center for Physics, the Santa Fe Institute, and the Department of Applied Mathematics and Theoretical Physics, University of Cambridge. We are grateful for the hospitality of these institutions. The work of JBH was supported in part by NSF grant PHY85-06686 and by a John Simon Guggenheim Fellowship. The work of MG-M was supported in part by the U.S. Department of Energy under contract DE-AC-03-81ER40050 and by the Alfred P. Sloan Foundation.

# NOTES

a. As in the "no boundary" and the "tunneling from nothing proposals" where the wave function of the universe is constructed from the action by a Euclidean functional integral in the first case or by boundary conditions on the implied Wheeler-DeWitt equation in the second. See, e.g., Refs 1 and 2.

b. The original paper is by Everett[9]. The idea was developed by many, among them Wheeler[10], DeWitt[11], Geroch[12], and Mukhanov[13] and independently arrived at by others e.g. Gell-Mann[14] and Cooper and VanVechten[15]. There is a useful collection of early papers on the subject in Ref 16.

c. It has been suggested[23] that, for application to highly quantum-mechanical spacetime, as in the very early universe, quantum mechanics should be generalized to yield a framework in which both time orderings are treated simultaneously in the sum-over-histories approach. This involves including both $\exp(iS)$ and $\exp(-iS)$ for each history and has as a consequence an evolution equation (the Wheeler-DeWitt equation) that is second order in the time variable. The suggestion is that the two time orderings decohere when the universe is large and spacetime classical, so that the usual framework with just one ordering is recovered.

d. An example of this occurs in the case of "null measurements" discussed by Renninger[34], Dicke[35], and others. An atom decays at the center of a spherical cavity. A detector which covers all but a small opening in the sphere does *not* register. We conclude that we have measured the direction of the decay photon to an accuracy set by the solid angle subtended by the opening. Certainly there is an interaction of the electromagnetic field with the detector, but did the escaping photon suffer an "irreversible act of amplification"? The point in the present approach is that the *set* of alternatives, detected and not detected, exhibits decoherence because of the place of the detector in the universe.

e. Perhaps as W. Unruh has suggested, there are complex adaptive systems, making no use of prediction, that can function in a highly quantum-mechanical way. If this is the case, they are very different from anything we know or understand.

# REFERENCES

For a subject as large as this one it would be an enormous task to cite the literature in any historically complete way. We have attempted to cite only papers that we feel will be *directly* useful to the points raised in the text. These are not always the earliest nor are they always the latest. In particular we have not attempted to review or to cite papers where similar problems are discussed from different points of view.